\documentclass[AMA,STIX1COL]{WileyNJD-v2}

\usepackage{placeins}
\usepackage{graphicx}
\usepackage{amsmath}
\usepackage{amssymb}
\usepackage{amsfonts}
\usepackage{subfig}
\usepackage{listings}

\articletype{Article Type}%

\received{}
\revised{}
\accepted{}
        
\begin{document}

\title{Task-based, GPU-accelerated and Robust Library for Solving Dense Nonsymmetric Eigenvalue Problems}

\author[1]{Mirko Myllykoski*}
\author[1]{Carl Christian Kjelgaard Mikkelsen}

\authormark{Mirko Myllykoski and Carl Christian Kjelgaard Mikkelsen}

\address[1]{\orgdiv{Department of Computing Science and HPC2N}, \orgname{Ume{\aa} University}, \orgaddress{\state{SE-901 87 Ume{\aa}}, \country{Sweden}}}

\corres{*Mirko Myllykoski, Department of Computing Science and HPC2N, Ume{\aa} University, SE-901 87 Ume{\aa}, Sweden. \email{mirkom@cs.umu.se}}

%\presentaddress{This is sample for present address text this is sample for present address text}

\abstract[Summary]{
In this paper, we present the StarNEig library for solving dense nonsymmetric standard and generalized eigenvalue problems. 
The library is built on top of the StarPU runtime system and targets both shared and distributed memory machines. 
Some components of the library have support for GPU acceleration.
The library is currently in an early beta state and supports only real matrices.
Support for complex matrices is planned for a future release.
This paper is aimed at potential users of the library.
We describe the design choices and capabilities of the library, and contrast them to existing software such as ScaLAPACK.
StarNEig implements a ScaLAPACK compatibility layer which should assist new users in the transition to StarNEig.
We demonstrate the performance of the library with a sample of computational experiments.
}


\maketitle

\section{Introduction}

In this paper, we present the StarNEig library \cite{starneig} for solving dense nonsymmetric standard and generalized eigenvalue problems.
StarNEig differs from existing libraries such as LAPACK \cite{lapack} and ScaLAPACK \cite{scalapack} by relying on a modern task-based approach (see, e.g., \cite{thibault} and references therein) in a manner similar to the already well-established PLASMA library\cite{10.1007/s10766-016-0441-6}.
Specifically, StarNEig is built on top of the StarPU runtime system \cite{starpu}.
This allows StarNEig to target both shared memory and distributed memory machines. 
Furthermore, some components of StarNEig have support for GPU acceleration.
The library is currently in an early beta state and under continuous development.
Currently, StarNEig applies to real matrices with real and complex eigenvalues and all calculations are done using real arithmetic.
This paper is an extended version of a conference paper by the authors\cite{ppam2019mm}.

This paper is addressed to potential users of the StarNEig library.
We hope that readers, who are already familiar with ScaLAPACK, will be able to decide if StarNEig is suitable for them.
In particular, we wish to communicate the changes needed to integrate existing ScaLAPACK style (or LAPACK style) software with StarNEig.
Central to this integration is the ScaLAPACK compatibility layer implemented in StarNEig.
This compatibility layer allows users to keep their existing two-dimensional block cyclic distribution of the data and call StarNEig routines directly to perform the computations. 
The authors hope to start a discussion which will help guide and prioritize the future development of the library.

Dense nonsymmetric eigenvalue problems are usually solved using three phases: reduction to condensed Hessenberg form, reduction to (real) Schur form and computation of eigenvectors.
Additionally, a fourth phase, called eigenvalue reordering, can be performed to acquire an invariant subspace that is associated with a given subset of eigenvalues.
All four phases are implemented in StarNEig in a task-based manner.
Performance-wise, the Hessenberg reduction phase in StarNEig is comparable to the LAPACK, ScaLAPACK and MAGMA\cite{magma} libraries, while the Schur reduction and the reordering phases in StarNEig are significantly faster than the ScaLAPACK implementations. 
Moreover, StarNEig can compute the eigenvectors directly from any Schur form without suffering from floating-point overflow, i.e., the implementation in robust. 
This functionality simply does not exist in ScaLAPACK and the implementation in StarNEig is significantly faster than the LAPACK implementation in a parallel setting.
We refer the reader to \cite{D27} for more comprehensive performance and accuracy evaluations.

The rest of this paper is organized as follows: 
Section \ref{sec:eigenproblem} provides a brief summary of the solution of dense nonsymmetric eigenvalue problems using the four phases mentioned earlier. 
Section \ref{sec:task-based} introduces the task-based approach and explains why the task-based approach can potentially lead to superior performance when compared to older, well-established techniques.
Section \ref{sec:starneig} introduces the reader to some of the inner workings of StarNEig.
In particular, the current state of the library and various limitations are explained in this section.
Section \ref{sec:performance} presents a sample of computational results which demonstrate the expected performance of StarNEig in both shared and distributed memory.
Finally, Section \ref{sec:summary} concludes the paper.

\section{Solution of dense nonsymmetric eigenvalue problems} \label{sec:eigenproblem}

Given a matrix $A \in \mathbb{R}^{n \times n}$, the standard eigenvalue problem consists of computing eigenvalues $\lambda_i \in \mathbb{C}$ and matching eigenvectors $x_i \in \mathbb{C}^n$, $x_i \not = 0$, such that
\begin{align}
 A x_i = \lambda_i x_i.\label{eq:standard}
\end{align}
Similarly, given matrices $A \in \mathbb{R}^{n \times n}$ and $B \in \mathbb{R}^{n \times n}$ the generalized eigenvalue problem for the matrix pair $(A,B)$ consists of computing generalized eigenvalues $\lambda_i \in \mathbb{C} \cup \{\infty\}$ and matching generalized eigenvectors $x_i \in \mathbb{C}^n$, $x_i \not = 0$, such that
\begin{align}
 A x_i = \lambda_i B x_i.\label{eq:generalized}
\end{align}

If the matrices $A$ and $B$ are sparse, then the well-known SLEPc library \cite{slepc} is one the better tools for solving the eigenvalue problems (\ref{eq:standard}) and (\ref{eq:generalized}).
Similarly, if the matrices $A$ and $B$ are is symmetric, then algorithms and software that take advantage of the symmetry are preferred (see, e.g., \cite{10.1007/s10766-016-0441-6,buttari2009class,Hiroto_Imachi2016,luszczek2011two,Marek_2014}).
Otherwise, if the matrices are both \emph{dense} and \emph{nonsymmetric}, then the route of acquiring the (generalized) eigenvalues and the (generalized) eigenvectors usually includes the following three phases:
\begin{description}

 \item[Hessenberg(-triangular) reduction:] The matrix $A$ or the matrix pair $(A,B)$ is reduced to Hessenberg form $H$ or Hessenberg-triangular form $(H,R)$ by a similarity transformation
 \begin{align}
  A = Q_1 H Q_1^T \; \text{ or } \; (A,B) = Q_1 (H,R) Z_1^T, \label{eq:hessenberg}
 \end{align}
 where $H$ is upper Hessenberg, $R$ is a upper triangular, and $Q_1$ and $Z_1$ are orthogonal matrices.
 
 \item[Schur reduction:] The Hessenberg matrix $H$ or the Hessenberg-triangular matrix pair $(H,R)$ is reduced to real Schur form $S$ or generalized real Schur form $(S,T)$ by a similarity transformation
 \begin{align}
  H = Q_2 S Q_2^T \; \text{ or } \; (H,R) = Q_2 (S,T) Z_2^T,
 \end{align}
 where $S$ is upper quasi-triangular with $1 \times 1$ and $2 \times 2$ blocks on the diagonal, $T$ is a upper triangular, and $Q_2$ and $Z_2$ are orthogonal matrices.
 The eigenvalues or generalized eigenvalues can be determined from the diagonal blocks of $S$ or $(S,T)$. 
% In particular, the $2 \times 2$ blocks on the diagonal of $S$ correspond to the complex conjugate pairs of (generalized) eigenvalues. 
 
 \item[Eigenvectors:] Finally, we solve for vectors $y_i \in \mathbb{C}^n$ from
 \begin{align}
  (S - \lambda_i I) y_i = 0 \; \text{ or } \; (S - \lambda_i T) y_i = 0 \label{eq:evec_solve}
 \end{align}
 and backtransform to the original basis by
 \begin{align}
  x_i = Q_1 Q_2 y_i \; \text{ or } \; x_i = Z_1 Z_2 y_i. \label{eq:evec_backtransform}
 \end{align}
 
\end{description}
Additionally, a fourth phase can be performed to acquire an invariant subspace of $A$ or $(A,B)$ that is associated with a given subset of eigenvalues or a given subset of generalized eigenvalues:
\begin{description}
 \item[Eigenvalue reordering:] The real Schur form $S$ or the generalized real Schur form $(S,T)$ is reordered, such that a selected set of eigenvalues or generalized eigenvalues appears in the leading diagonal blocks of an updated real Schur form $\hat S$ or an updated generalized real Schur form $(\hat S, \hat T)$, by a similarity transformation
 \begin{align}
  S = Q_3 \hat S Q_3^T \; \text{ or } \; (S,T) = Q_3 (\hat S, \hat T) Z_3^T,
 \end{align}
 where $Q_3$ and $Z_3$ are orthogonal matrices.
\end{description}
See \cite{Golub1996} for a detailed explanation of the underlying mathematical theory.

\section{A case for the task-based approach} \label{sec:task-based}

A task-based algorithm functions by cutting the computational work into self-contained tasks that all have a well defined set of inputs and outputs.
In particular, StarNEig divides the matrices into disjoint (square) tiles and each task takes a set of tiles as its input and produces/modifies a set of tiles as its output.
The main difference between tasks and regular function/subroutine calls is that a task-based algorithm does not call the associated computation kernels directly.
Instead, the tasks are inserted into a runtime system which then derives the data dependences between the tasks from the supplied input and output information.
The runtime system then schedules the tasks to computational resources, such as CPUs and GPUs, in a sequentially consistent order as dictated by the data dependences.
The main benefit from this is that as long as the cutting is carefully done, the underlying parallelism is exposed automatically as the runtime system gradually traverses the resulting task graph.
In particular, the runtime system can detect and tap into previously unexplored avenues of parallelism that are hidden within the task graphs.
This leads to significantly more powerful algorithms that are able to adapt to different inputs and changing hardware configurations.
Other benefits of the task-based approach include, for example, better load balancing and resource utilization due to dynamic scheduling, task priorities and implicit MPI communications that are automatically derived from the task graph.

The following subsections briefly discuss the four phases in the algorithm stack from the point of view of task parallelism.
We do not attempt to explain the details of each algorithm, rather we focus on the key steps and explain how they benefit from task parallelism.
We will use the standard eigenvalue problem as an illustration.

\subsection{GPU-accelerated Hessenberg reduction}

\begin{figure}[h]
\centering
\subfloat[Panels.\label{fig:reduce_a}]{
 \includegraphics[width=0.21\textwidth]{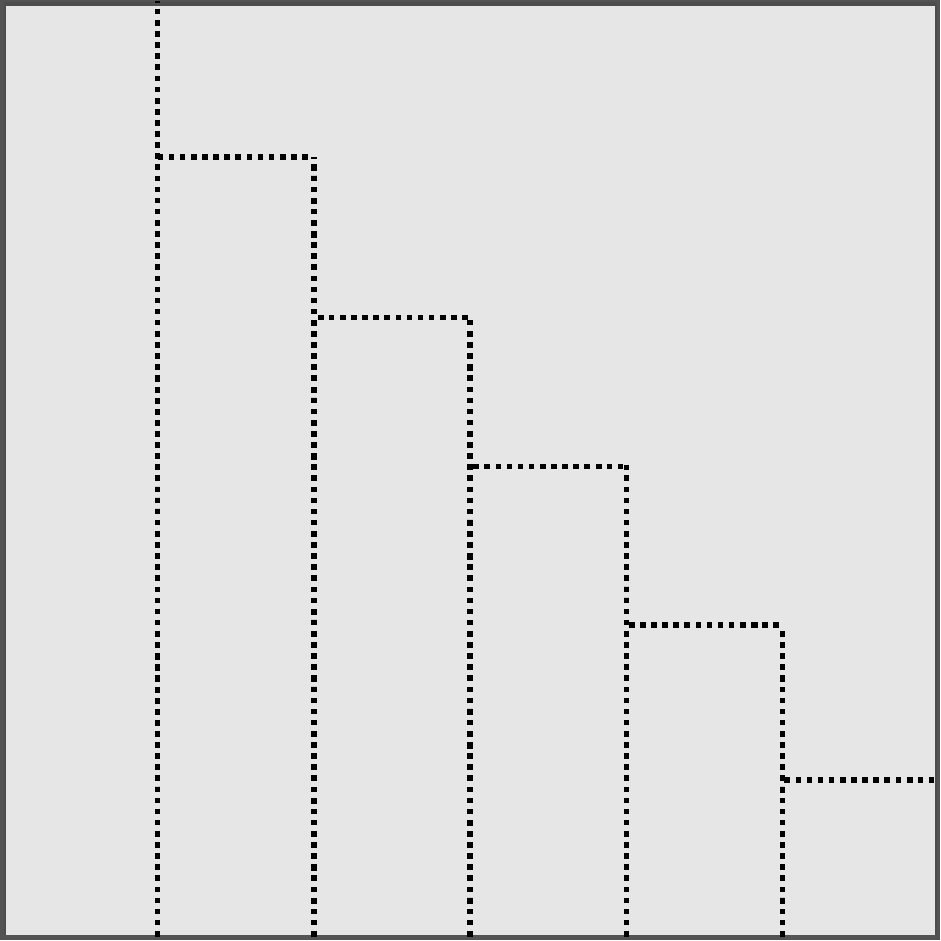}
}
\;
\subfloat[Reduce the panel.\label{fig:reduce_b}]{
 \includegraphics[width=0.21\textwidth]{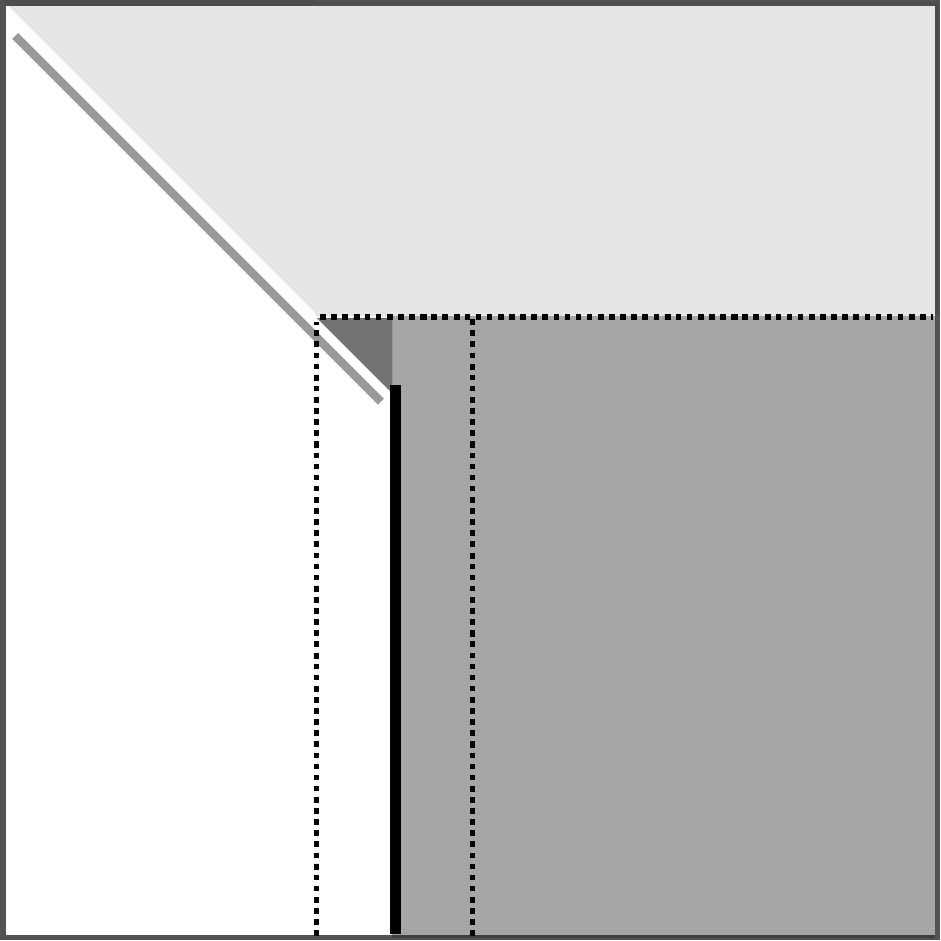}
}
\;
\subfloat[Update the trailing matrix.\label{fig:reduce_c}]{
 \includegraphics[width=0.21\textwidth]{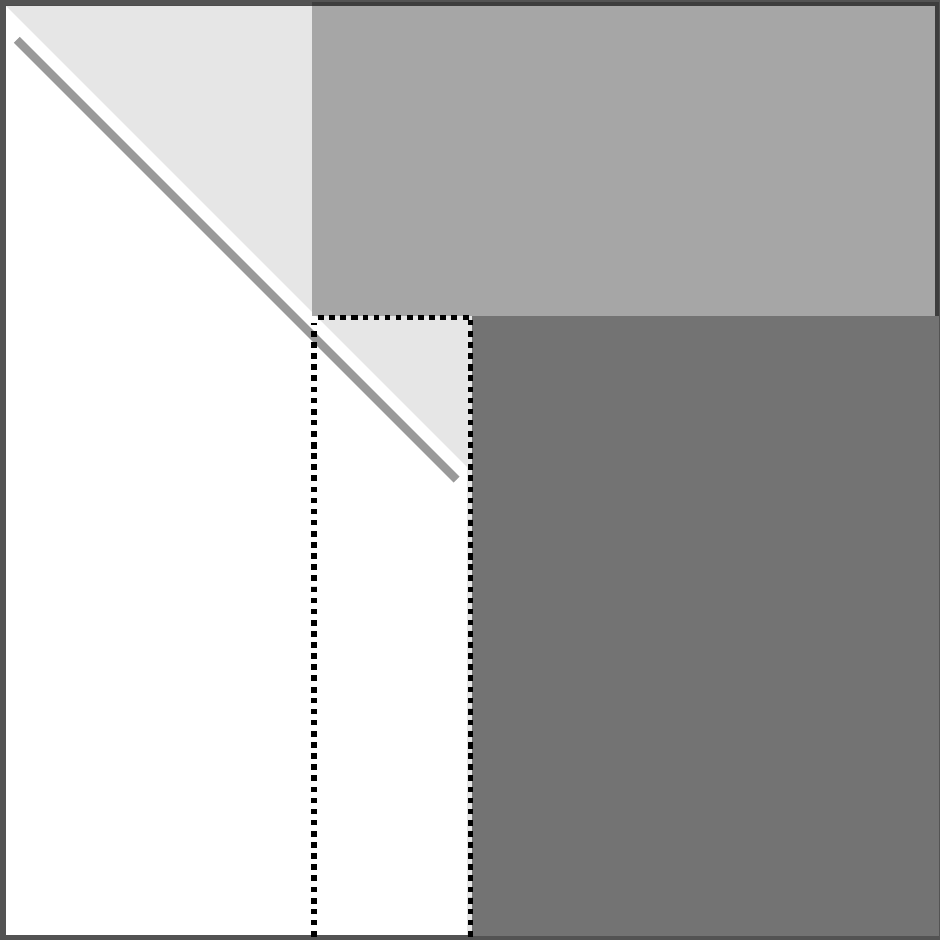}
}
\;
\subfloat[Update the top part from the right.\label{fig:reduce_d}]{
 \includegraphics[width=0.21\textwidth]{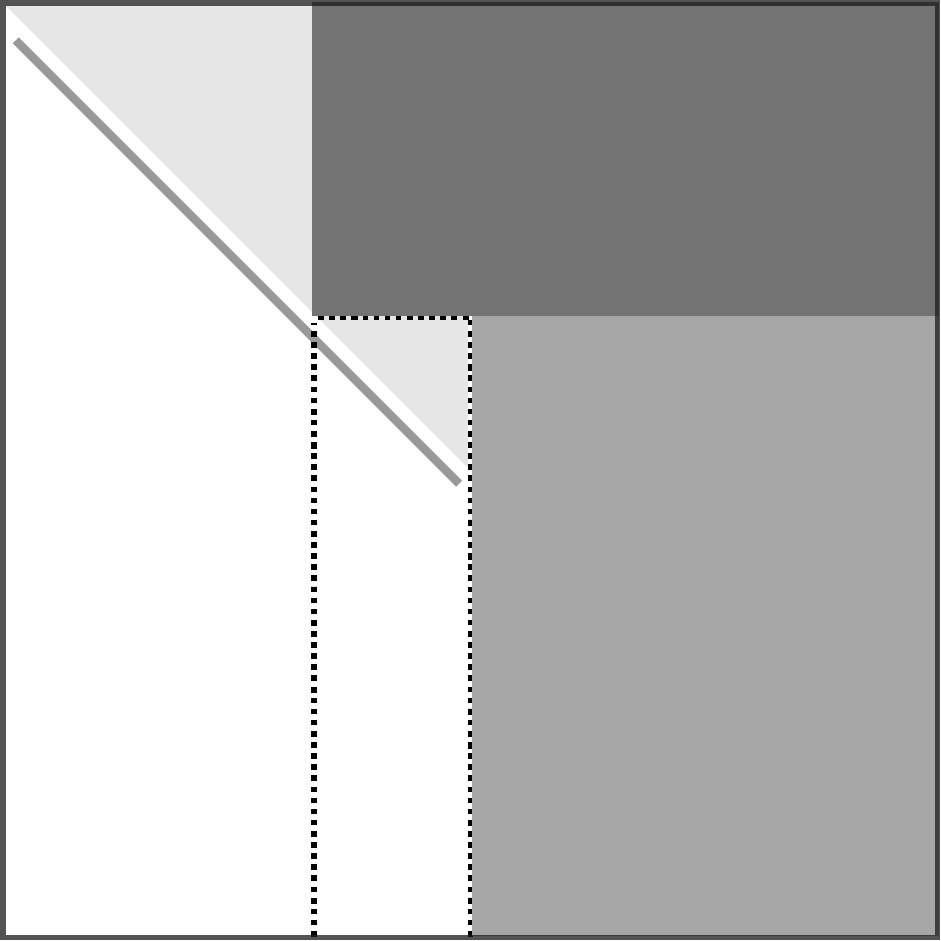}
}
 \caption{
 An illustration of the standard algorithm for reducing a matrix to upper Hessenberg form.
 }
 \label{fig:reduce}
\end{figure}

We begin by discussing the Hessenberg reduction phase.
We emphasize that this phase does not benefit as much from the task-based approach at the other phases.
% Even though this phase does not benefit from the task-based approach as much the other phases, we still cover it for the sake of completeness.
Now, the so-called standard algorithm \cite{Quintana-Orti2006} for reducing a nonsymmetric matrix $A$ to upper Hessenberg form $H$ first divides the matrix $A$ into disjoint panels as illustrated in Figure \ref{fig:reduce_a}.
Each panel is then reduced to upper Hessenberg form as summarized below:
\begin{enumerate}
 \item \emph{Reduce the panel.}
 The $i$th column in the panel is reduced by constructing and applying a suitable Householder reflector $I - \tau_i v_i v_i^T$, where $\tau_i \in \mathbb{R}$ and $v_i \in \mathbb{R^n}$.
 All involved reflectors are initially applied only inside the current panel and accumulated into a compact WY representation $I - V_i T_i V_i^{T}$, where $T_i \in \mathbb{R}^{i \times i}$ is upper triangular and $V_i = [v_1 v_2 \dots v_i]$.
 In tandem, one of the necessary intermediate results is also accumulated into a matrix $Y_i = A V_i T_i$.
 The construction of the matrix $Y_i$ allows us to reduce the entire panel without updating the other sections of the matrix. 
 However, the update formula includes a large matrix-vector multiplication involving the section of the matrix that trails the current column (see the shaded area in Figure \ref{fig:reduce_b}) and the non-zero part of the vector $v_i$.
 Although these matrix-vector multiplications constitute approximately only 20\% of the total number of flops, they are significantly more expensive than the remaining 80\% due to the fact that matrix-vector multiplication is a memory bound operation.
 This is an important factor to consider when analysing the performance.
  
\item \emph{Update the trailing matrix.}
 Update the section of the matrix that trails the current panel (see the shaded area in Figure \ref{fig:reduce_c}) by the update formula
 \begin{align}
  A \gets (I - V T V^{T})^{T} (A - Y V^{T}),
 \end{align}
 where $I - V T V^{T}$ and $Y$ are the final compact WY representation and the final intermediate result matrix from the panel reduction step, respectively.
  
\item \emph{Update the top part from the right.}
 Update the section of the matrix above the current panel (see the shaded area in Figure \ref{fig:reduce_d}) by the update formula
 \begin{align}
  A \gets A - Y V^{T}.
 \end{align}
  
\end{enumerate}

\begin{figure}[h]
 \centering
 \includegraphics[width=0.9\textwidth]{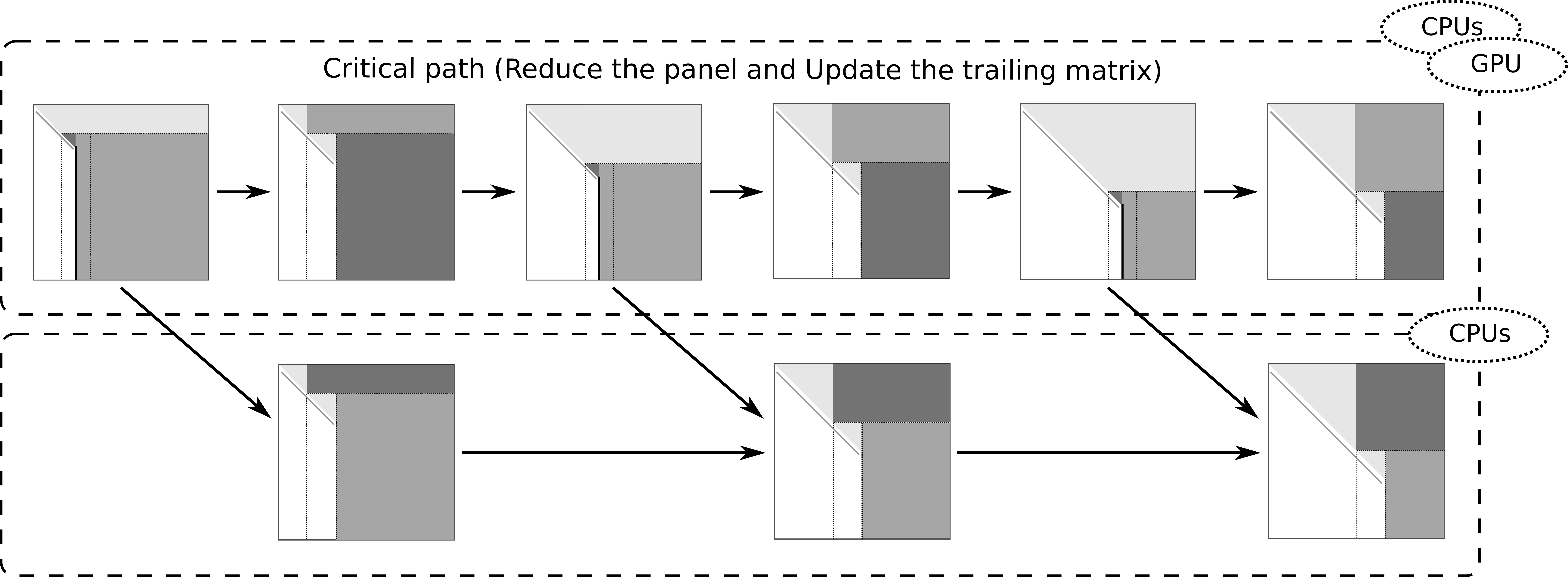}
 \caption{
 An illustration of the two scheduling contexts.
 }
 \label{fig:reduce2}
\end{figure}

In StarNEig, each one of the aforementioned three steps is formulated as a task.
In particular, for task scheduling overhead related reasons, the entire \textit{reduce the panel} step is implemented as one monolithic task.
Implementing each column reduction as a set of separate tasks would lead to an excessive number of lightweight tasks and thus to an unmanageable amount of scheduling overhead.
The computational resources are divided into two scheduling contexts:
\begin{description}

 \item[Parallel scheduling context] contains a subset of the available CPU cores and a GPU, if one exists in the system. 
 Each inserted task is executed in parallel by all included CPU cores or by the GPU, depending on which resource is predicted to be the best option by the runtime system.
 In order to accomplish this, the runtime system uses calibrated performance models to predict the execution and data transfer times for each task.
 The \textit{reduce the panel} and \textit{update the trailing matrix} tasks form the critical path of the algorithm and are scheduled to this scheduling context as illustrated in Figure \ref{fig:reduce2}.
 The CPU implementation of the \textit{reduce the panel} task copies the trailing matrix to memory buffers that are allocated from the local NUMA islands.
 That is, each involved CPU core has its own local memory buffer and the obtainable memory bandwidth is thus significantly higher compared to a situation where the CPU cores would access the memory across the NUMA islands.
 Note that the trailing matrix is copied only once in the beginning of each \textit{reduce the panel} task.
 This is an another reason for implementing \textit{reduce the panel} step as one monolithic task.
 
\item[Sequential scheduling context] contains the remaining CPU cores and will inherit computational resources from the parallel scheduling context once the critical path has been completed.
Each inserted task is executed sequentially by one of the available computational resources.
 The \textit{update the top part from the right} tasks are scheduled to this context as illustrated in Figure \ref{fig:reduce2}.
 Note that the \textit{update the top part from the right} tasks never feeds back into the critical path and can therefore be scheduled independently.
 The tasks that compute the matrix $Q_1$ in (\ref{eq:hessenberg}) are also scheduled to this context and given lower priority.
 
\end{description}

The main benefit that comes from the task-based approach here is that the runtime system is allowed to schedule the work to the GPU when it predicts this will improve the performance.
In particular, the runtime system usually schedules \textit{reduce the panel} tasks to the GPU because most modern GPUs have a much higher memory bandwidth than CPUs.
This means that the time that is spent computing the large matrix-vector multiplications is significantly reduced.
The runtime system will handle the necessary data transfer between main memory and GPU memory, including prefetching of the data when necessary.
The end result is that the task-based implementation will naturally behave very similarly to the implementation available in MAGMA library \cite{magma,tomov2009accelerating} but provides some additional flexibility as the scheduling decisions are done dynamically.

\subsection{Aggressive Early Deflation, Bulge Chasing and Eigenvalue Reordering}

\begin{figure}[h]
 \centering
 \includegraphics[width=0.90\textwidth]{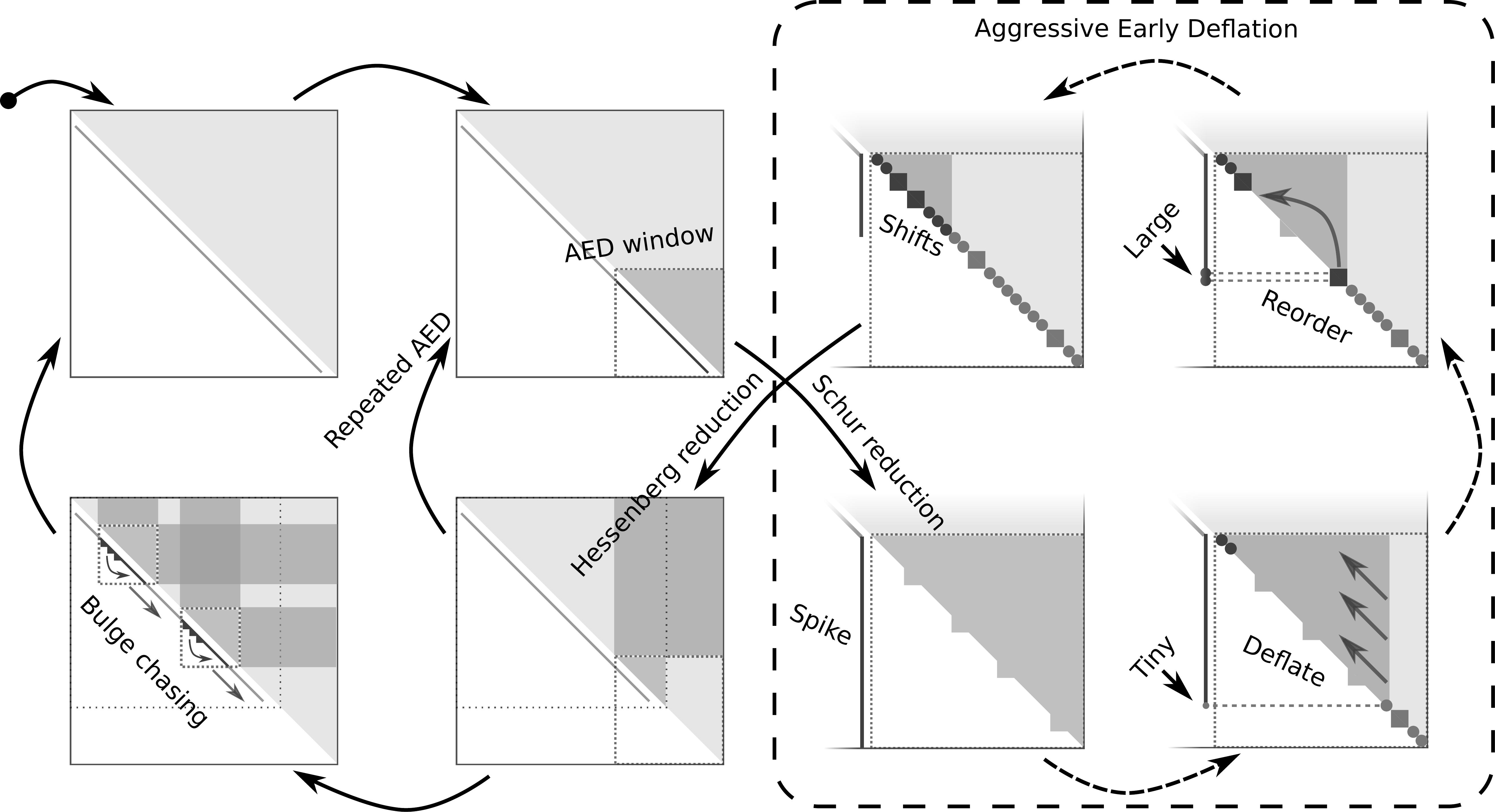}
 \caption{
 An illustration of the multi-shift QR algorithm with aggressive early deflation.
 }
 \label{fig:qr}
\end{figure}

We will now use the Schur reduction and eigenvalue reordering phases to illustrate some of the more notable benefits of the task-based approach.
The modern approach for obtaining a Schur form $S$ of a nonsymmetric matrix $A$ is to apply the multi-shift QR algorithm with Aggressive Early Deflation (AED) to the upper Hessenberg form $H$ (see \cite{Braman2002,Braman2002a,GraKagKreShao2015a,GraKagKreShao2015b} and references therein).
The algorithm is a sequence of steps of two types, AED and bulge chasing, as illustrated in Figure \ref{fig:qr} and summarized below:
\begin{description}

 \item [AED step] first reduces a small diagonal window (a.k.a AED window) to real Schur form via a recursive application of the QR algorithm.
 When the sections of the matrix outside the AED window are updated from the left, a spike is induced to the left of the AED window and we will thus temporarily deviate from the upper Hessenberg form.
 All diagonal blocks in the reduced AED window are then systematically evaluated in order to identify those eigenvalues that can be safely deflated without introducing significant perturbations.
 % If the corresponding row in the spike is found to be small enough (i.e., so-called deflation condition is satisfied), then the element in the spike is set to zero and the eigenvalue that corresponds to the diagonal block is deflated.
  If the corresponding element in the spike is found to be small enough (i.e., the so-called deflation condition is satisfied), then the element is set to zero and the eigenvalue that corresponds to the diagonal block is deflated.
 On the other hand, if the corresponding entry in the spike is found to be too large, then the remaining AED window is reordered such that the diagonal block that failed the deflation check is moved to the upper left corner of the window thus pushing the remaining unevaluated diagonal blocks downwards along the diagonal.
 After all diagonal blocks have been evaluated, the remaining spike is eliminated by performing a small-sized Hessenberg reduction.
 
\item[Bulge chasing step] chases a set of $3 \times 3$ bulges down the diagonal.
 The eigenvalues of the diagonal blocks that failed the deflation condition, $\lambda_1, \lambda_2, \dots, \lambda_m$, are used as shifts to generate the bulges.
 That is, the first column of the matrix $H$ is transformed to the first column of the matrix $(H - \lambda_1 I)(H - \lambda_2 I)$ via a small-sized Householder reflector.
 When applied from the both sides, the reflector creates fill-in in the form a $3 \times 3$ bulge that appears in the upper left corner of the matrix.
 At this point, the bulge could be eliminated by chasing it down the diagonal with a sequence of overlapping small-sized Householder reflectors and this would complete one implicit QR iteration.
 However, a multi-shift QR algorithm will instead chase the bulge just enough so that a second bulge can be introduced using the shifts $\lambda_3$ and $\lambda_4$.
 The same procedure is then repeated until a total of $m/2$ bulges have been introduced.
 The bulges are then chased in groups down the diagonal to complete one pipelined QR iteration.
 The bulge chasing step is then followed by a second AED step and the same procedure is repeated until all eigenvalues have been deflated or an iteration limit is reached.
 
\end{description}
Similarly, the eigenvalue reordering phase is based on applying sequences of overlapping Givens rotations and small-sized Householder reflectors to $S$.
The Schur form $S$ is essentially reordered in a bubble sort manner by using kernels that swap two adjacent diagonal blocks \cite{Bai1993a}.

\begin{figure}[h]
\centering
\subfloat[Individual scalar updates.\label{fig:differences_a}]{
 \includegraphics[width=0.24\textwidth]{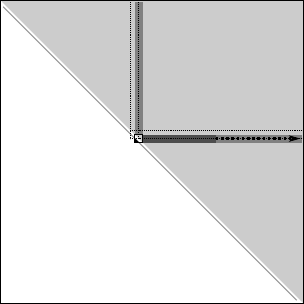}
}
\;
\subfloat[Localized updates inside a window.\label{fig:differences_b}]{
 \includegraphics[width=0.24\textwidth]{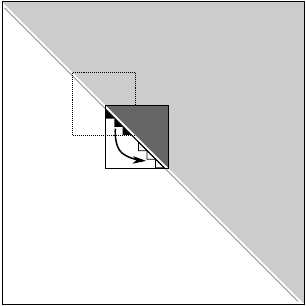}
}
\;
\subfloat[BLAS level-3 off-diagonal updates.\label{fig:differences_c}]{
 \includegraphics[width=0.24\textwidth]{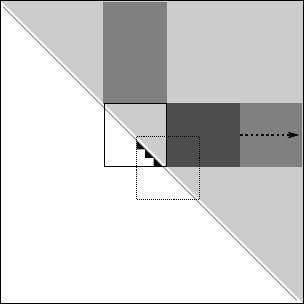}
}

\subfloat[Concurrent windows in ScaLAPACK.\label{fig:differences_d}]{
 \includegraphics[width=0.24\textwidth]{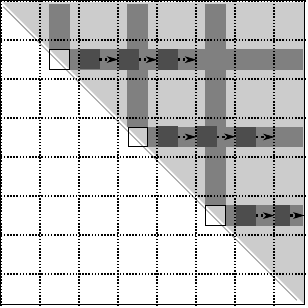}
}
\;
\subfloat[Concurrent windows in StarNEig.\label{fig:differences_e}]{
 \includegraphics[width=0.24\textwidth]{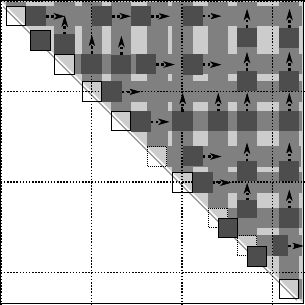}
}
\caption{ 
 Hypothetical snapshots taken during the computations.
 The currently active regions are highlighted with darker shade and the propagation directions of the off-diagonal updates are marked with arrows.
 In (a), the overlap between two overlapping transformations is highlighted with dashed lines.
 In (b) and (c), the overlap between two diagonal windows is highlighted with dashed lines.
 In (d) and (e), the dashed lines illustrate how the matrix is divided into distributed blocks.
}\label{fig:differences}
\end{figure}

If the Givens rotations and small-sized Householder reflectors are applied one by one, then memory is accessed as shown in Figure \ref{fig:differences_a}.
This is grossly inefficient for two reasons: i) the transformations are so localized that parallelizing them would not produce any significant speedup and ii) the matrix elements are touched only once thus leading to very low arithmetic intensity.
The modern approach (see \cite{Braman2002,Braman2002a,GraKagKreShao2015a,GraKagKreShao2015b,Kressner2006c,Granat_2009} and references therein) groups together a set of local transformation and initially applies them to a relatively small diagonal window as shown in Figure \ref{fig:differences_b}.
The localized transformations are accumulated into an accumulator matrix and later propagated as BLAS level-3 operations acting on the off-diagonal sections of the matrix as shown in Figure \ref{fig:differences_c}.
This leads to much higher arithmetic intensity and enables proper parallel implementations as \textit{multiple} diagonal windows can be processed concurrently.
These are the main reason why the multi-shift QR algorithm introduces several $3 \times 3$ bulges to the diagonal.
The $3 \times 3$ bulges are divided into groups and each group is chased separately down the diagonal.
A set, or a \emph{chain}, of overlapping diagonal windows is associated with each group. 
Similarly, several selected diagonal blocks can be grouped and moved together in the eigenvalue reordering phase\cite{Kressner2006c,Granat_2009,Myllykoski2018}.

The Schur reduction and eigenvalue reordering phases are implemented in ScaLAPACK as {\tt PDHSEQR} \cite{GraKagKreShao2015a} and {\tt PDTRSEN} \cite{Granat_2009} subroutines, respectively.
Following the ScaLAPACK convention, the matrices are distributed in a two-dimensional block cyclic fashion \cite{blacs}.
The resulting memory access pattern is illustrated in Figure \ref{fig:differences_d} for a $3 \times 3$ MPI process mesh.
In this example, three diagonal windows can be processed simultaneously. 
The related BLAS level-3 off-diagonal updates require careful coordination since the left and right hand side updates must be performed in a sequentially consistent order.
In practice, this means (global or row/column communicator broadcast) synchronization after each set of BLAS level-3 off-diagonal updates have been applied.
In addition, each AED step introduces a global synchronization point.

In StarNEig, the Schur reduction and eigenvalue reordering phases are implemented with the following tasks types:
\begin{description}
 \item[Window task] generates and applies a set of local transformations inside a diagonal window. 
 Takes the intersecting tiles as input, and produces updated tiles and an accumulator matrix as output. 
 \item[Right update task] applies accumulated right-hand side updates using BLAS level-3 operations. 
 Takes the intersecting tiles and an accumulator matrix as input, and produces updated tiles as output.
 \item[Left update task] applies accumulated left-hand side updates using BLAS level-3 operations. 
 Takes the intersecting tiles and an accumulator matrix as input, and produces updated tiles as output.
\end{description}

\begin{figure}[h]
 \centering
 \includegraphics[scale=0.5]{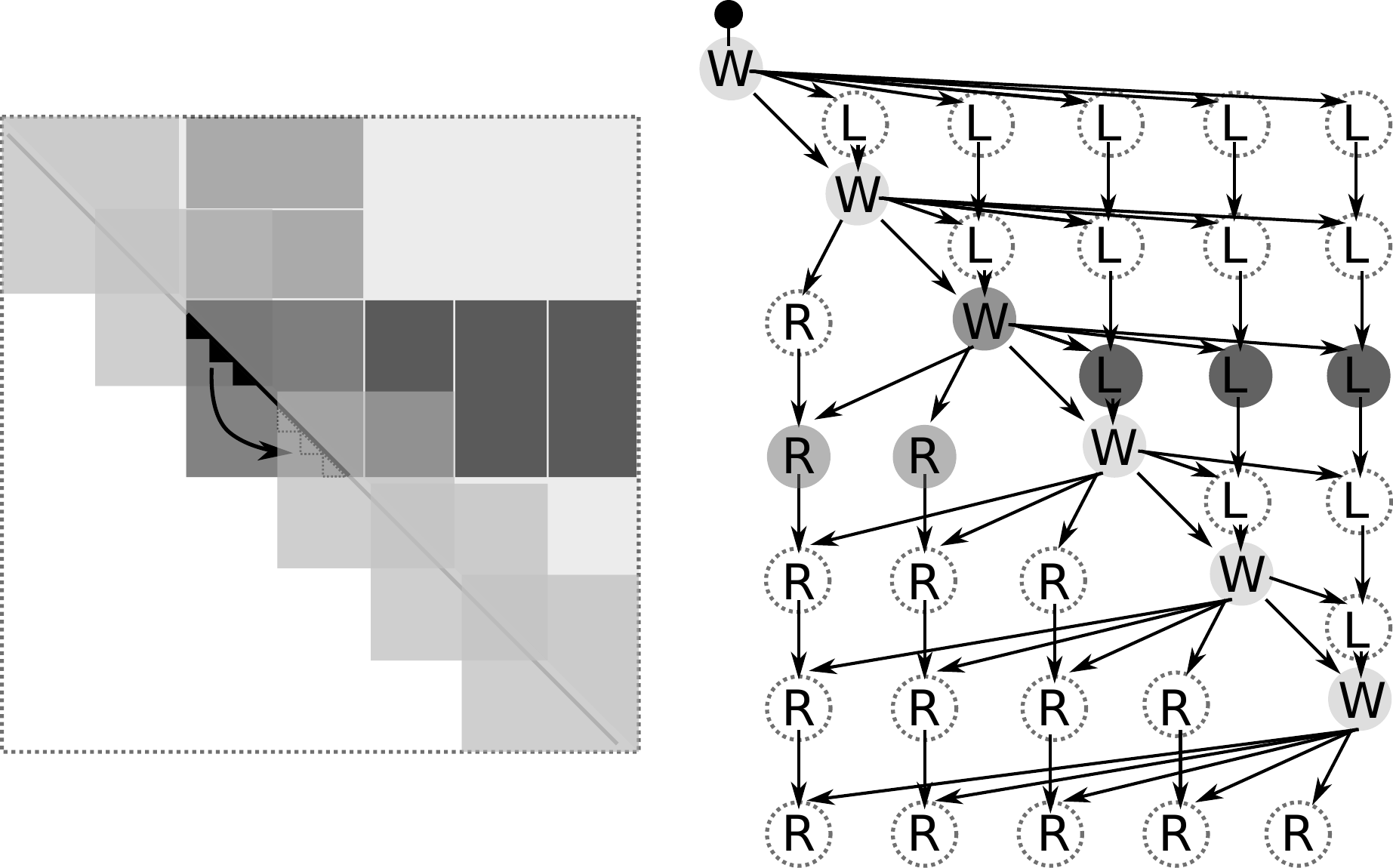}
 \caption{
 A hypothetical task graph arising from a situation where a set of three bulges is chased down the diagonal.
 We have simplified the graph by omitting dependences between the left (L) and right (R) update tasks as these dependences are enforced through the diagonal bulge chasing tasks (W).
 Note that the actual bulge chasing step involves \emph{several} sets of bulges and the resulting task graph is therefore significantly more complex than the simplified graph presented here.
 }
 \label{fig:chain_flow}
\end{figure}

\begin{figure}[p]
\centering
\subfloat[The first bulge chasing step with delayed right-hand side updates.\label{fig:frame128}]{
 \includegraphics[width=0.40\textwidth]{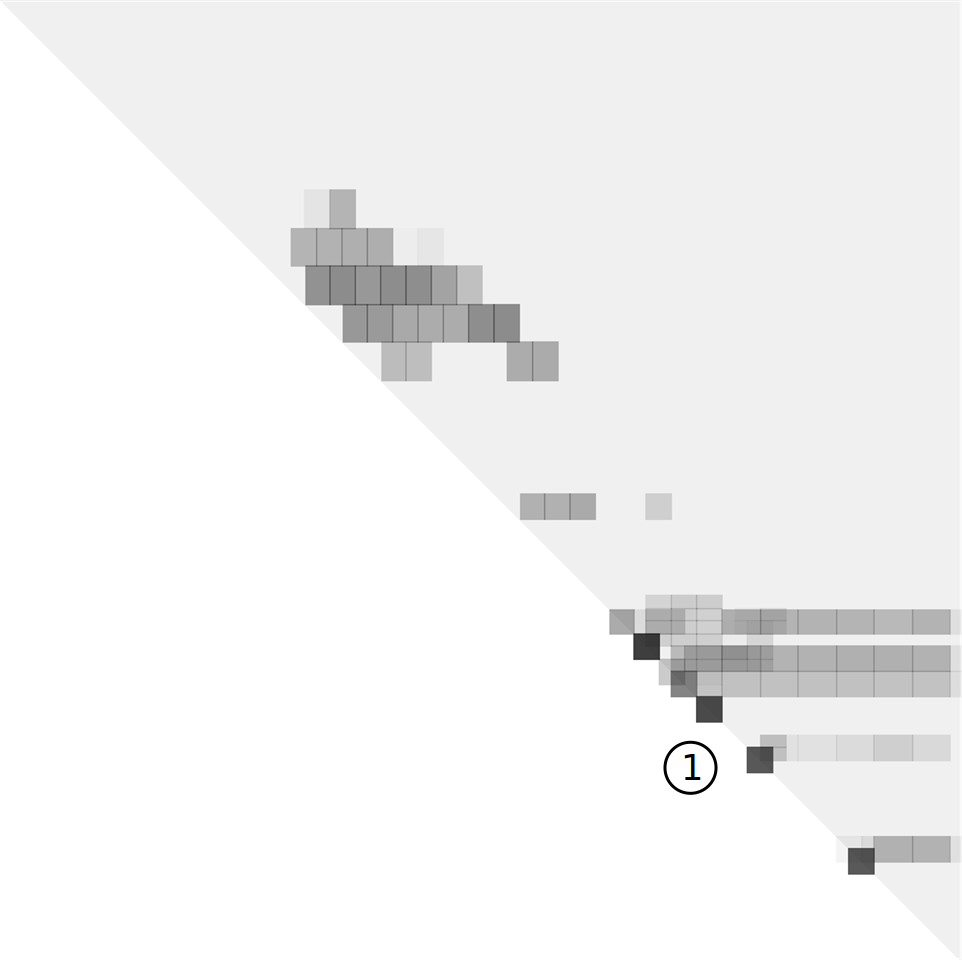}
}
\;\;\;
\subfloat[Overlapping AED and bulge chasing steps.\label{fig:frame254}]{
 \includegraphics[width=0.40\textwidth]{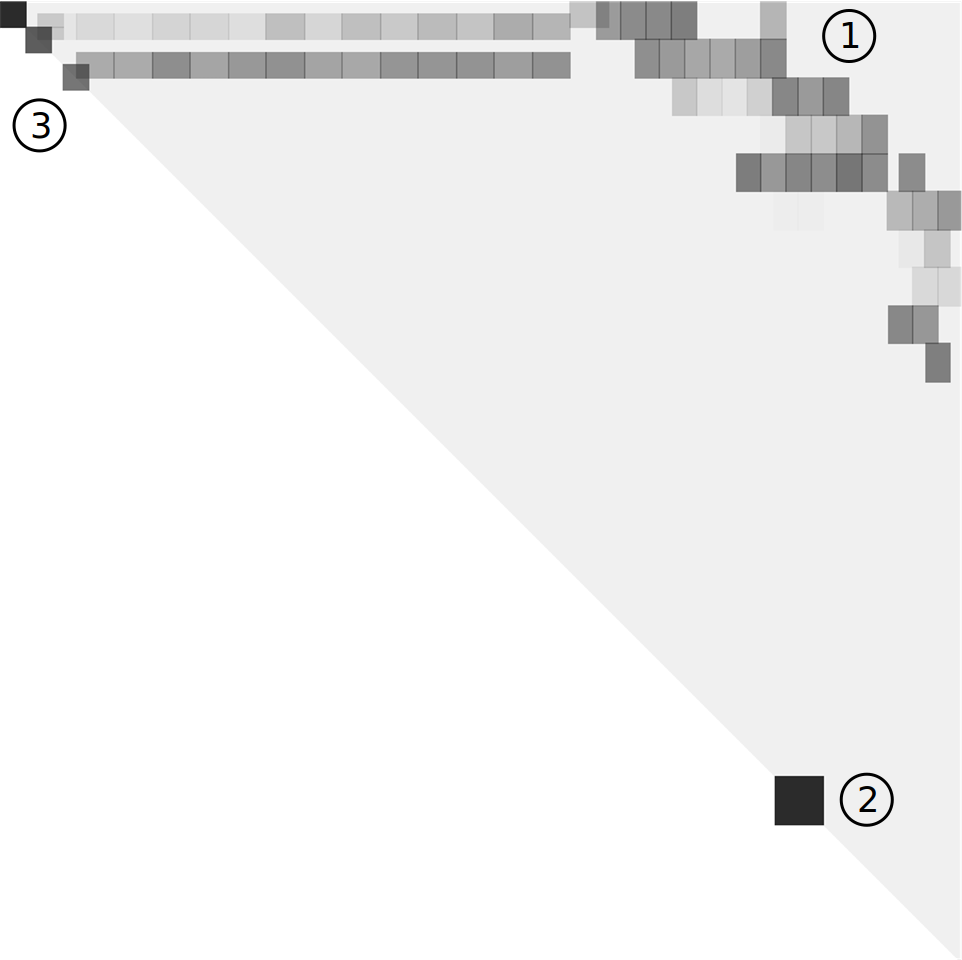}
}
\caption{ 
 Snapshots taken during the first two iterations of the multi-shift QR algorithm with AED.
 The numbering matches the numbering in Figure \ref{fig:qr2}. 
 That is, (1) is the bulge chasing step from the first iteration, (2) is the small-sized Hessenberg reduction from the second AED step and (3) is the bulge chasing step from the second iteration.
 Note that the three steps are merged in (b).
}\label{fig:frames}
\end{figure}

\begin{figure}[p]
 \centering
 \includegraphics[width=0.64\textwidth]{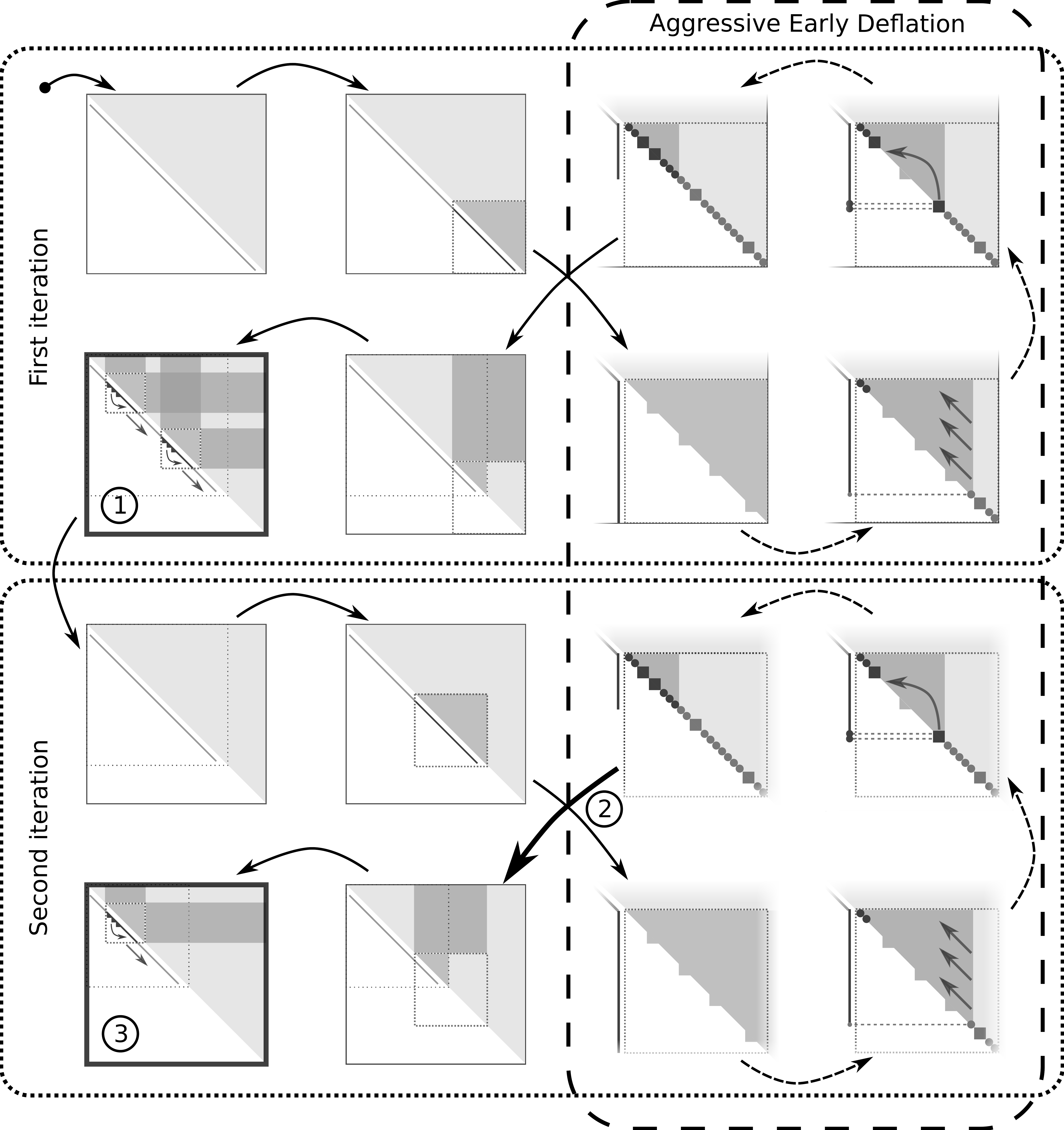}
 \caption{
 An illustration of the first two iterations of the multi-shift QR algorithm with AED.
 The steps that are concurrently active in Figure \ref{fig:frame254} are highlighted in bold and numbered correspondingly.
 }
 \label{fig:qr2}
\end{figure}

The tasks are inserted into the runtime system in a sequentially consistent order and each chain of overlapping diagonal windows leads to a task graph like the one shown in Figure \ref{fig:chain_flow}.
Note that real live task graphs are significantly more complex than shown here, but also enclose more opportunities for parallelism. 
It is also critical to realize that the runtime system \textit{guarantees} that the tasks are executed in a sequentially consistent order.
In particular, there is no need for synchronization and different computational steps are allowed to overlap (see Figure~\ref{fig:differences_e}) as the runtime system \emph{merges} the corresponding sub-graphs together.
This can lead to a much higher concurrency since idle time can be reduced by delaying low priority tasks until computational resources start becoming idle.
This is exactly what can be observed from Figure \ref{fig:frame128} where a subset of the right-hand side updates are given lower priority and are therefore scheduled only after the diagonal bulge chasing tasks and the left-hand side update tasks cannot saturate all available computational resources.
This is possible because,  as seen in Figure \ref{fig:chain_flow}, most right-hand side update tasks do not feed back to the other sections of the task graph and can therefore be scheduled independently.

The AED step can also be overlapped with the bulge chasing steps as shown in Figures \ref{fig:frame254} and \ref{fig:qr2}.
This improves the concurrency significantly compared to ScaLAPACK because the ScaLAPACK implementation will effectively synchronize before and after each AED step.
This means that in ScaLAPACK most MPI ranks will become idle while a subset of the MPI ranks are involved in  the AED step.
Actually, StarNEig can even overlap two bulge chasing steps with each other as seen in Figures \ref{fig:frame254}.
See \cite{D27,D26} for further information.

\subsection{Robust computation of eigenvectors}

Mathematically, the problem of computing a single eigenvector of, say, a quasi-triangular matrix is trivial.
The standard algorithm is a variant of substitution.
However, substitution is very vulnerable to floating-point overflow.
In particular, there exist triangular linear systems which are well-conditioned in the sense of Skeel for which the solution grows exponentially and rapidly exceeds the representational range of any floating-point number system\cite{ccpe2018}.
We say that an algorithm or a subroutine is robust if all intermediate and final results are in the representational range, i.e., floating point overflow is avoided.
In LAPACK there exists a robust family {\tt xLATRS} of subroutines for solving triangular linear system $Tx=b$\cite{lawn36}.
They dynamically scale the entire right-hand side and return a scaling factor $\alpha$ and a vector $x$ such that $Tx = \alpha b$.
The purpose of the scaling factor $\alpha$ is to extend the floating-point representational range.
In LAPACK, the solvers for computing eigenvectors from Schur forms are all descended from {\tt xLATRS}.
They are scalar codes which compute the eigenvectors one by one.
The back-transformation to the original basis can of course be done using BLAS level-3 operations.
In ScaLAPACK, there is no parallel implementation of {\tt xLATRS} and the solvers for computing eigenvectors are not robust.

In contrast, StarNEig implements novel algorithms for computing eigenvectors which are tiled, parallel and robust. In StarNEig, each matrix of eigenvectors is partitioned into tiles $X_{ij}$.
Every tile $X = \begin{bmatrix} x_1 & x_2 & \cdots & x_k \end{bmatrix}$ is augmented with a vector of scaling factors $\alpha \in \mathbb{R}^k$ with one scaling factor per column.
The augmented tile $\langle \alpha,X \rangle$ represents the matrix $Y = \begin{bmatrix} y_1 & y_2 & \cdots & y_k \end{bmatrix} $ given by $y_j = x_j/\alpha_j$.
The StarNEig solvers for computing eigenvectors accept and produce augmented tiles.
This allows StarNEig to obtain a representation of the eigenvectors without exceeding the representational range.
A final post-processing step ensures that all segments of each eigenvector are consistently scaled.
In reality, StarNEig is exploiting a principle which is familiar to every scientist: Two results can be combined if we know how to convert between the different system of measurements, say, the metric system and the imperial system.

The use of augmented tiles is compatible with the linear update $Z \gets Y - TX $ and the vast majority of the arithmetic operations can be completed using BLAS level-3 operations.
Given a matrix $T$ and two augmented tiles $\langle\alpha, X\rangle$ and $\langle\beta, Y\rangle$, such that $Y - TX$ is defined, StarNEig produces an augmented tile $\langle\gamma,Z\rangle$ which represents the intended result, i.e., 
\begin{equation}
  \gamma_j^{-1} z_j = \beta_j^{-1} y_j - \alpha_j^{-1} T x_j.
\end{equation}
This is achieved as follows.
A preprocessing step ensures that each pair $(x_j,y_j)$ of columns of $X$ and $Y$ are not only consistently scaled, but the linear update $z_j = y_j - Tx_j$ can be computed without exceeding the overflow threshold.
If $X$, $Y$ and $T$ are $n \times n$ matrices, then this preprocessing step requires $O(n^2)$ operations.
Then the linear update $Z \gets Y - TX$ is executed using a single BLAS level-3 operation.
We emphasize that the cost of the preprocessing step is insignificant compared with $O(n^3)$ arithmetic operations required for the linear update.
Rescaling vectors to obtain a consistent scaling requires that the scaling factors are nonzero.
Otherwise, a division-by-zero is attempted.
In StarNEig, the possibility of the scaling factors underflowing is significantly reduced by using scaling factors which are integer powers of $2$.
Currently StarNEig uses at least 32 bit signed integers, for which the smallest scaling factor is $\alpha = 2^{-2^{31}+1} \approx 10^{-6.4646 \times 10^{6}}$, but this can easily be extended to the point where underflow is a practical impossibility using 64 bit unsigned integers.

The main benefits that stem from the task-based approach are the merging of different computational steps and the improved load balancing.
Additional information can be found in the existing literature. 
In particular, the fundamental principles for solving triangular linear systems in parallel without suffering from overflow are discussed in \cite{ppam2017cckm,ccpe2018cckm}.
The StarNeig solver for computing generalized eigenvectors is the subject of a separate paper \cite{ppam2019cckm}.

\section{StarNEig library} \label{sec:starneig}

StarNEig is a C-library that runs on top of the StarPU task-based runtime system.
StarPU handles low-level operations such as heterogeneous scheduling, data transfers and replication between memory spaces and MPI communication between compute nodes.
In particular, StarPU is responsible for managing the various computational resources such as CPU cores and GPUs.
The support for GPUs and distributed memory were the main reasons why StarPU was chosen as the runtime system.

StarPU manages a set of worker threads; usually one thread per computational resource.
In addition, one thread is responsible for inserting the tasks into StarPU and tracking the state of the machine. 
If necessary, one additional thread is allocated for MPI communication.
For these reasons, StarNEig must be used in a \textit{one process per node} (1ppn) configuration, i.e., several CPU cores should be allocated for each MPI process (a node can be a full node, a NUMA island or some other reasonably large collection of CPU cores).

\begin{table}[h]
\caption{
 Current status of the StarNEig library.
 }
\label{tab:starneig_status}
\centering
\begin{tabular}{l|ccc}
\textbf{Computational phase}         & \textbf{Shared memory} & \textbf{Distributed memory} & \textbf{GPUs (CUDA)} \\ \hline
\textbf{Hessenberg reduction}   & Complete               & ScaLAPACK wrapper                   & Single GPU supported    \\
\textbf{Schur reduction}        & Complete               & Complete                    & Experimental  \\
\textbf{Eigenvalue reordering}  & Complete               & Complete                    & Experimental  \\
\textbf{Eigenvectors} & Complete               & In progress                 & ---  \\ \hline
\textbf{Hessenberg-triangular reduction}       & LAPACK wrapper     & ScaLAPACK wrapper                   & ---   \\
\textbf{Generalized Schur reduction} & Complete   & Complete                    & Experimental  \\
\textbf{Generalized eigenvalue reordering}      & Complete   & Complete                    & Experimental  \\
\textbf{Generalized eigenvectors}    & Complete   & In progress                 & ---
\end{tabular}
\end{table}

The current status of StarNEig is summarized in Table \ref{tab:starneig_status}.
The library is currently in an early beta state.
The \emph{Experimental} status indicates that the software component has not been tested as extensively as those software components that are considered \emph{Complete}.
In particular, the GPU functionality requires some additional involvement from the user (performance model calibration).
At the time of writing this paper, only real arithmetic is supported and certain interface functions are implemented as LAPACK and ScaLAPACK wrapper functions.
However, we emphasize that StarNEig supports real valued matrices that have complex eigenvalues and eigenvectors.
Additional distributed memory functionality and support for complex data types are planned for a future release.

\subsection{Distributed memory}

\begin{figure}[h]
\centering
\subfloat[Distributed blocks.\label{fig:distr_matrix1}]{
 \includegraphics[width=0.25\textwidth]{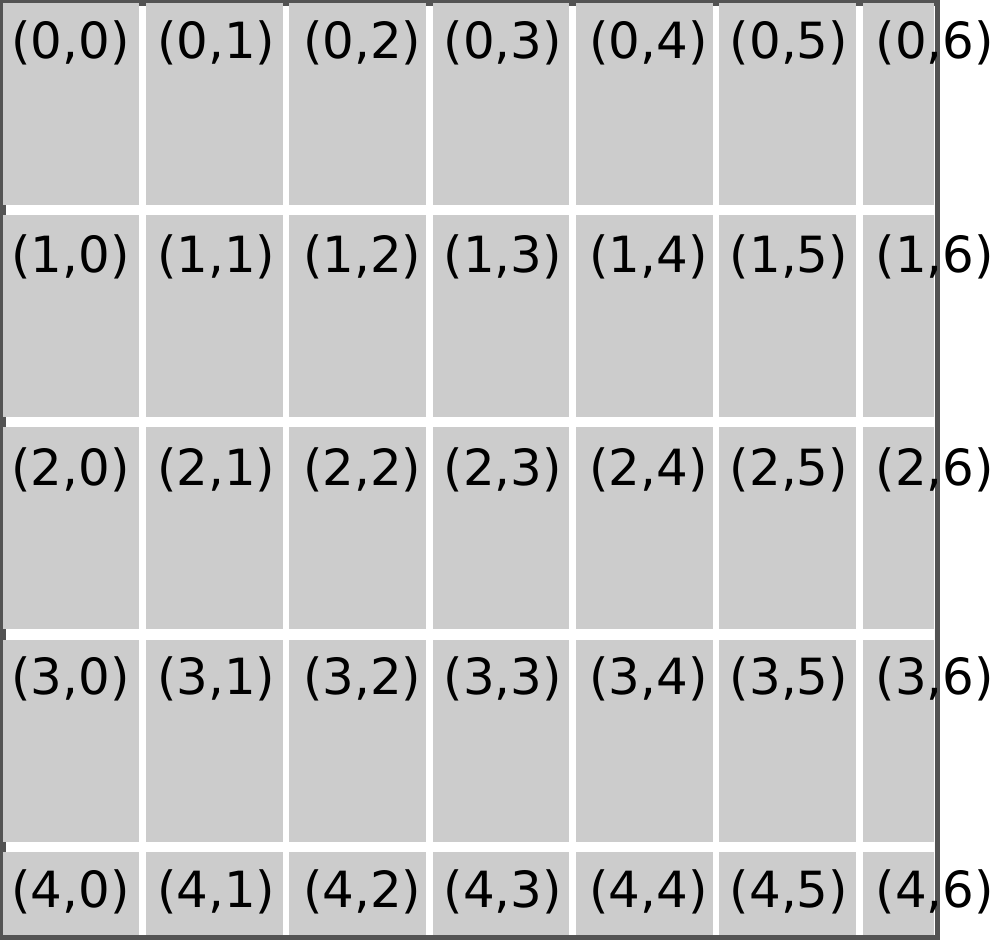}
}
\subfloat[Arbitrary distribution.\label{fig:distr_matrix2}]{
 \includegraphics[width=0.25\textwidth]{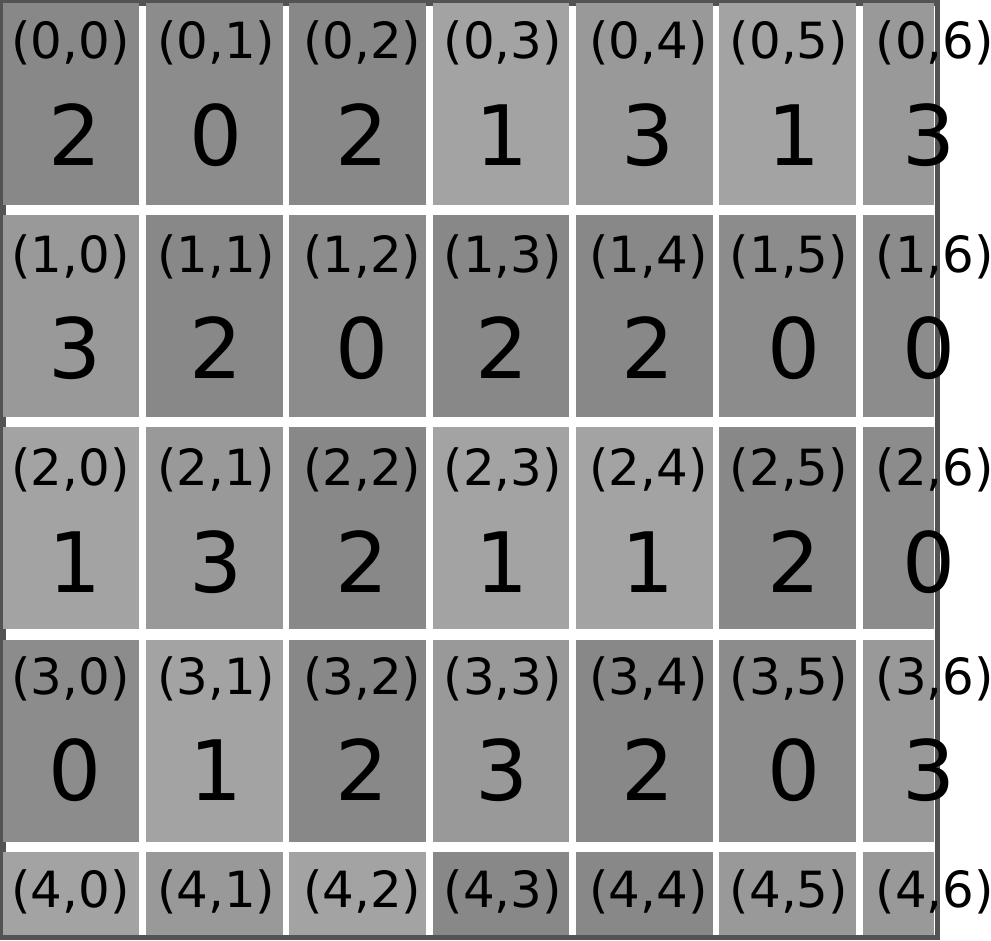}
}
\subfloat[2D-BCD.\label{fig:distr_matrix3}]{
 \includegraphics[width=0.25\textwidth]{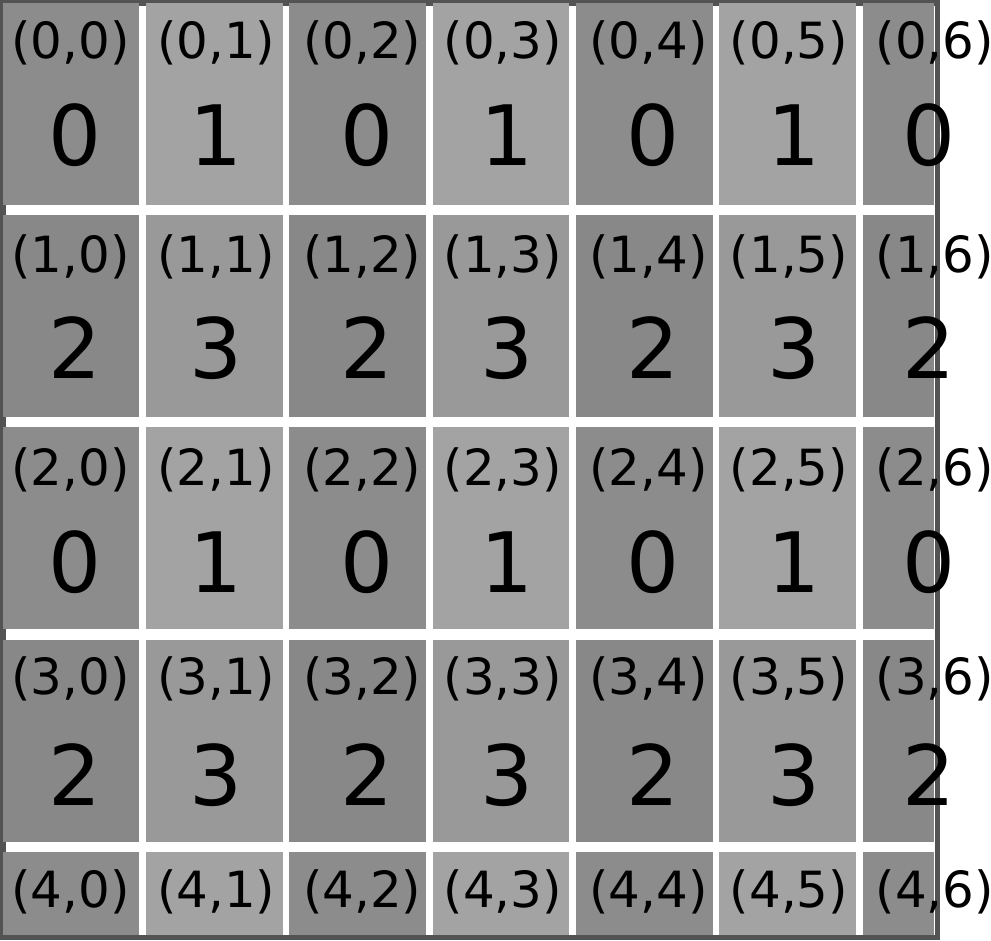}
}
\caption{Examples of various data distributions supported by StarNEig, including two-dimensional block cyclic distribution (2D-BCD).}\label{fig:distr_matrix}
\end{figure}

StarNEig distributes the matrices in rectangular blocks of uniform size (excluding the last block row and column) as illustrated in Figure \ref{fig:distr_matrix1}.
The data distribution, i.e., the mapping from the distributed blocks to the MPI process rank space, can be arbitrary as illustrated in Figure \ref{fig:distr_matrix2}.
A user has three options:
\begin{enumerate}
 \item Use the default data distribution. 
 This is recommended for most users and leads to reasonable performance in most situations.
 \item Use a two-dimensional block cyclic distribution (see Figure \ref{fig:distr_matrix3}). In this case, the user may select the MPI process mesh dimensions and the rank ordering. 
 \item Define a data distribution function $d : \mathbb{Z^+} \times \mathbb{Z^+} \to \mathbb{Z^+}$ that maps the block row and column indices to the MPI rank space.
 For example, in Figure \ref{fig:distr_matrix2}, rank $0$ owns the blocks (0,1), (1,2), (1,5), (1,6), (2,6), (3,0) and (3,5).
\end{enumerate}
The library implements distribution agnostic copy, scatter and gather operations.

Users who are familiar with ScaLAPACK are likely accustomed to using relatively small distributed block sizes (between 64--256).
In contrast, StarNEig functions optimally only if the distributed blocks are relatively large (at least 1000 but preferably must larger).
This is due to the fact that StarNEig further divides the distributed blocks into tiles and a tiny tile size leads to excessive task scheduling overhead because the tile size is closely connected to the task granularity.
Furthermore, as mentioned in the preceding section, StarNEig should be used in 1ppn configuration as opposed to a \textit{one process per core} (1ppc) configuration which is common with ScaLAPACK.

\subsection{ScaLAPACK compatibility}

StarNEig is fully compatible with ScaLAPACK and provides a ScaLAPACK compatibility layer that encapsulates BLACS contexts and descriptors \cite{blacs} inside transparent objects, and implements a set of bidirectional conversion functions.
The conversions are performed in-place and do not modify any of the underlying data structures.
Users can mix StarNEig interface functions with ScaLAPACK routines without intermediate conversions.
The use of the ScaLAPACK compatibility layer requires the use of either the default data distribution or the two-dimensional block cyclic data distribution.
Listing~\ref{code:scalapack} illustrates how a distributed StarNEig matrix is converted to an equivalent BLACS context and a matching local buffer.
The matrix is then reduces to upper Hessenberg form by calling the \texttt{PDGEHRD} routine from ScaLAPACK.
This is actually how the ScaLAPACK wrapper functions are implemented in StarNEig, see Table~\ref{tab:starneig_status}.

\lstset{emph={ 
    dA, distr, ctx, ptr, descr
    },emphstyle={\bfseries\underbar}
}

\begin{lstlisting}[caption={An example how a distributed matrix is converted to a BLACS descriptor and a local buffer.},label=code:scalapack,float,language=C]
// create a 2D block cyclic data distribution (pm x pn process mesh)
starneig_distr_t distr =
    starneig_distr_init_mesh(pm, pn, STARNEIG_ORDER_DEFAULT);

// create a n x n distributed matrix (bn x bn blocks) 
starneig_distr_matrix_t dA =
    starneig_distr_matrix_create(n, n, bn, bn, STARNEIG_REAL_DOUBLE, distr);

...

// convert the data distribution to a BLACS context
starneig_blacs_context_t ctx = starneig_distr_to_blacs_context(distr);

// convert the distributed matrix to a BLACS descriptor and a local buffer
starneig_blacs_descr_t descr;
double *ptr;
starneig_distr_matrix_to_blacs_descr(dA, ctx, &descr, (void **)&ptr);

// ScaLAPACK routine for reducing a general distributed matrix to upper
// Hessenberg form
extern void pdgehrd_(int const *, int const *, int const *, double *,
    int const *, int const *, starneig_blacs_descr_t const *, double *,
    double *, int const *, int *);
    
pdgehrd_(&n, &ilo, &ihi, ptr, &ia, &ja, &descr, tau, ...);
\end{lstlisting}

\section{Performance evaluation} \label{sec:performance}

Computational experiments were performed on the Kebnekaise system, located at the High Performance Computing Center North (HPC2N), Ume{\aa} University.
Kebnekaise is a heterogeneous systems consisting of many different types of compute nodes.
The compute node types relevant for this paper are:
\begin{description}
 \item[Broadwell compute node] contains 28 Intel Xeon E5-2690v4 cores organized into 2 NUMA islands with 14 cores each and 128 GB memory. The nodes are connected with FDR Infiniband. All distributed memory experiments were performed on these nodes.
 \item[Skylake compute node] contains 28 Intel Xeon Gold 6132 cores organized into 2 NUMA islands with 14 cores each and 192 GB memory. All Hessenberg reduction experiments were performed on these nodes.
 \item[Large memory node] contains 72 Intel Xeon E7-8860v4 cores organized into 4 NUMA islands with 18 cores each and 3072 GB memory. All eigenvector experiments were performed on these nodes.
 \item[V100 GPU node] contains 28 Intel Xeon Gold 6132 cores organized into 2 NUMA islands with 14 cores each and 192 GB memory. Each NUMA island is connected to a single NVIDIA Tesla V100 GPU. All GPU experiments were performed on these nodes.
\end{description}

For the distributed memory and eigenvalue reordering related GPU experiments, the software was compiled with GCC 7.3.0 and linked to OpenMPI 3.1.3, OpenBLAS 0.3.2, ScaLAPACK 2.0.2, CUDA 9.2.88, and StarPU 1.2.8.
Since the the version of \texttt{PDHSEQR} routine that exists in ScaLAPACK 2.0.2 is known to be buggy, StarNEig was actually compared against an updated version of {\tt PDHSEQR}; see \cite{GraKagKreShao2015a}.
The updated version places far less strict conditions on the distributed block size and thus perform better on modern hardware.
For the Hessenberg reduction and eigenvector computation experiments, the software was compiled with ICC 19.0.1.144 and linked to Intel MPI 2018.4.274, Intel MKL 2019.1.144, CUDA 10.1.105 and StarPU 1.2.8.
StarPU was compiled with the \texttt{-{}-disable-cuda-memcpy-peer} configuration flag enabled.
This significantly reduces the CUDA related overhead.

In most experiments, StarNEig version 0.1-beta.2 was used.
However, the Schur reduction and eigenvalue reordering experiments in distributed memory were performed using an older and unpublished version of StarNEig.
The main difference between this unpublished version and the version 0.1-beta.2 is the deflation condition used in the Schur reduction phase.
The deflation condition is used to decide when an eigenvalue can be deflated and thus impacts the convergence rate of the algorithm.
The unpublished version uses a deflation condition that is identical to the one used in LAPACK and \texttt{PDHSEQR} where as the version 0.1-beta.2 uses the so-called norm stable deflation condition \cite{Braman2002,Braman2002a}.
The latter deflation condition is less strict and could thus potentially lead to faster convergence.
The latest version of StarNEig (0.1-beta.4) allows the user to choose between these two deflation conditions.

\begin{table}[h]
\centering
\caption{
 A run time comparison (in seconds) between parallel MKL-LAPACK, MKL-ScaLAPACK, MAGMA and StarNEig when computing a Hessenberg form in shared memory.
 Columns 2-4 show the execution times for 28 Intel Xeon Gold 6132 cores and columns 5-6 show the execution times for 14 Intel Xeon Gold 6132 cores paired with a NVIDIA Tesla V100 GPU. The \texttt{STARPU\_WORKERS\_CPUID} environmental variable was set to \texttt{0 14 1 15 2 16 }[...].
} 
\label{tab:sep_hessenberg_compare}
\begin{tabular}{r | c c c | c c}
        & \multicolumn{3}{c|}{28 cores} & \multicolumn{2}{c}{14 cores + V100 GPU} \\
$n$     & LAPACK & ScaLAPACK & StarNEig & MAGMA & StarNEig \\
\hline
 5\,000	& 3.8 & 4.2 & 8.4 & 1.3 & 2.2 \\
10\,000	& 32  & 29 & 40  & 7.3 & 8.9 \\
15\,000	& 113 & 95 & 117 & 23  & 24  \\
20\,000	& 278 & 223 & 249 & 49  & 53  \\
25\,000 & 561 & 459 & 454 & 90  & 97  \\
30\,000 & 953 & 728 & 755 & 144 & 164 \\
40\,000 & 2397 & 1722 & 1711 & --- & --- \\
\end{tabular}
\end{table}

Table \ref{tab:sep_hessenberg_compare} shows how the Hessenberg reduction routine in StarNEig compares against LAPACK (with parallel BLAS), ScaLAPACK (in single node) and MAGMA\cite{magma,tomov2009accelerating}.
Since the implementations in all four libraries are based on the standard algorithm, the performance is limited by the throughput of the matrix-vector multiplication routine.
It is therefore important that the memory access pattern is optimized.
In particular, a memory access to a NUMA island that is not local to a particular core is very likely to reduce the performance since the obtainable memory bandwidth is significantly lower compared to a local access.
A memory access pattern that is close to optimal is easy to achieve with ScaLAPACK since each MPI process accesses its own local buffer and this buffer can be allocated from the NUMA island that is closest to the core to which the MPI process is bound.
For the other CPU experiments, the memory was allocated in interleaved mode across the two NUMA islands in order to evenly distribute the load.
Since the matrix-vector multiplication operation is memory bound, it is not surprising that StarNEig closely matches the performance of LAPACK, ScaLAPACK and MAGMA.
It is only in the case of the smallest considered matrix where the additional task scheduling related overhead begins to negatively effect StarNEig's performance.
The additional memory locality considerations in StarNEig begin to show their effect when the matrices are reasonably large, leading up to 40\% performance improvement compared to LAPACK.

\begin{table}[h]
\centering
\caption{
 A run time comparison (in seconds) between ScaLAPACK and StarNEig in distributed memory.
 Each node contains 28 Intel Xeon E5-2690v4 cores.
 }
\label{tab:sep_schur_compare}
\begin{tabular}{r | c c | c c | c c}
                         & \multicolumn{2}{c|}{CPU cores} & \multicolumn{2}{c|}{Schur reduction} & \multicolumn{2}{c}{Eigenvalue reordering} \\
\multicolumn{1}{c|}{$n$} & \;ScaLAPACK\; & \;StarNEig\;     & \;ScaLAPACK\; & \;StarNEig\; & \;ScaLAPACK\; & \;StarNEig\; \\ \hline
10\,000 & 36 & 28 & 38 & 18 & 12 & 3 \\
20\,000 & 36 & 28 & 158 & 85 & 72 & 25 \\
40\,000 & 36 & 28 & 708 & 431 & 512 & 180 \\
60\,000 & 121 & 112 & 992 & 563 & 669 & 168 \\
80\,000 & 121 & 112 & 1667 & 904 & 1709 & 391 \\
100\,000 & 121 & 112 & 3319 & 1168 & 3285 & 737 \\
120\,000 & 256 & 252 & 3268 & 1111 & 2902 & 581 \\
\end{tabular}
\end{table}

Table \ref{tab:sep_schur_compare} shows a comparison between ScaLAPACK and StarNEig.
All distributed memory experiments were performed using a square MPI process grid.
This was done because the maximum number of diagonal bulge chasing and reordering windows in the \texttt{PDHSEQR} and \texttt{PDTRSEN} routines is given by $\min(pm,pn)$, where $pm$ and $pn$ are the height and width of the process mesh, respectively. 
A square process mess thus maximizes the degree of parallelism in the ScaLAPACK routines.
We always map each StarNEig process to a full node (28 cores) and each ScaLAPACK process to a single CPU core, the latter being done for the same reason as the choice to use a square process mesh.
Since it is not straightforward to find a CPU core allocation that would lead to a square MPI process grid in both configurations, the number of CPU cores in each ScaLAPACK experiment is always equal or larger than the number of CPU cores in the corresponding StarNEig experiment.
The upper Hessenberg matrices for the Schur reduction experiments were computed from random matrices (entries uniformly distributed over the interval $[-1,1]$).
In the ScaLAPACK experiments, the matrices were distributed in $160 \times 160$ blocks, and in the StarNEig experiments, the library default block size was used.
In the eigenvalue reordering experiments, 35\% of the diagonal blocks were randomly selected.
From Table \ref{tab:sep_schur_compare}, we note that StarNEig is between 1.6 and 2.9 times faster than ScaLAPACK ({\tt PDHSEQR}) when computing the Schur form and between 2.8 and 5.0 times faster than ScaLAPACK ({\tt PDTRSEN}) when reordering the Schur form.
The distributed memory experiments were initially reported in the technical report \cite{D27}.

\begin{figure}[h]
 \centering
 \includegraphics[scale=0.75]{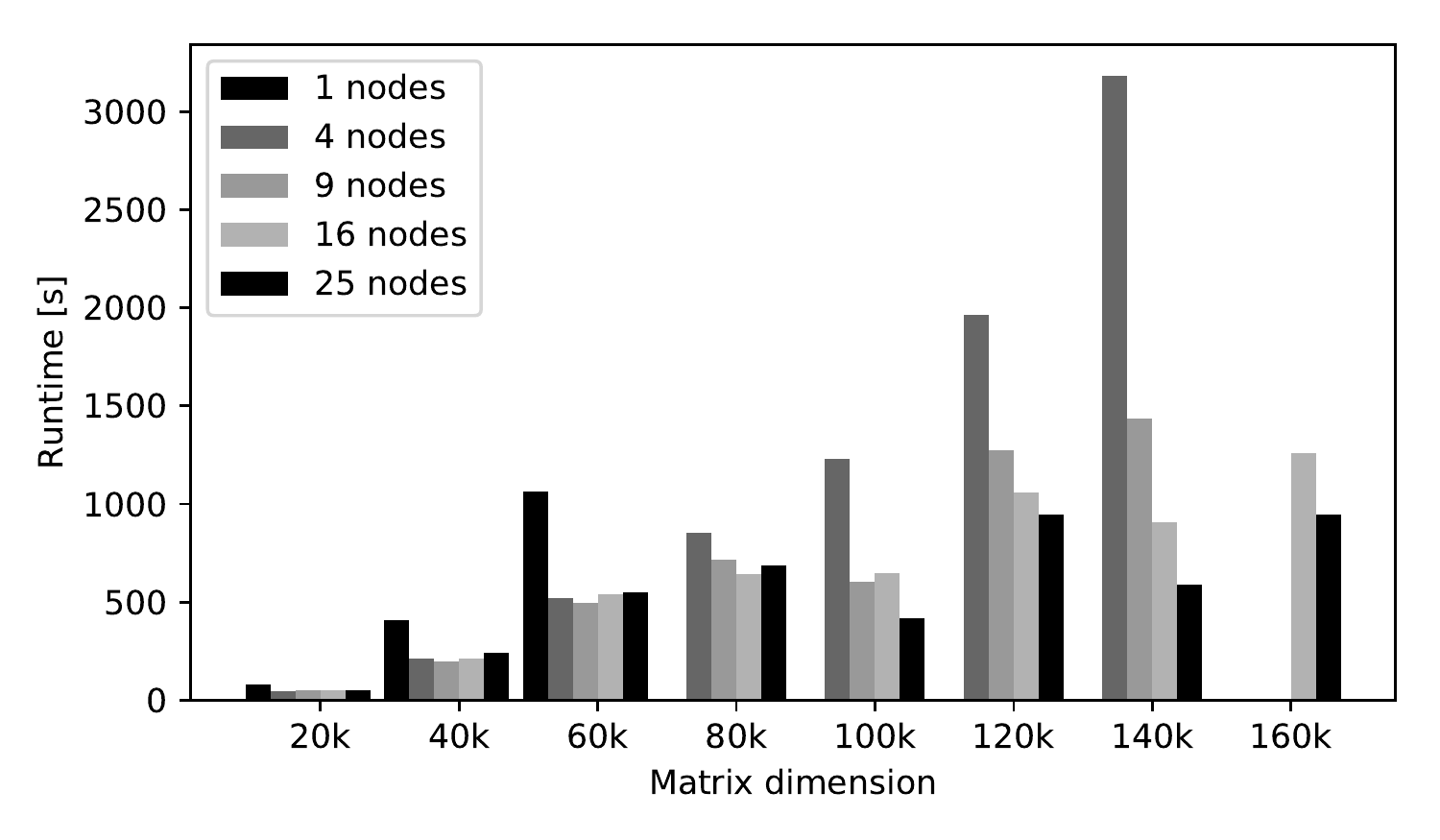}
 \caption{
 Distributed memory scalability of StarNEig when computing a Schur form.
 Each node contains 28 Intel Xeon E5-2690v4 cores. 
 }
 \label{fig:sep_schur_scalability}
\end{figure}

\begin{figure}[h]
 \centering
 \includegraphics[scale=0.75]{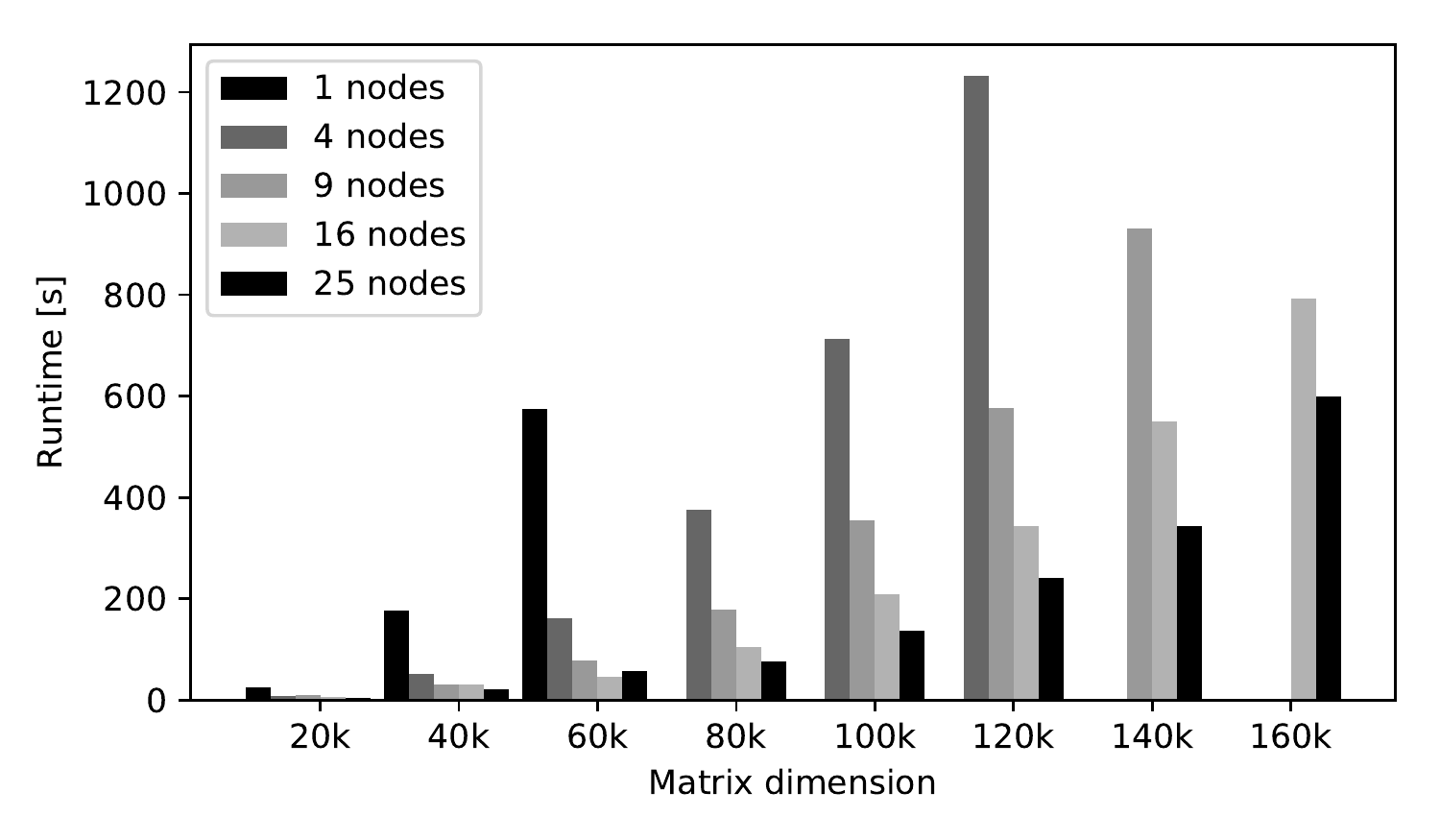}
 \caption{
 Distributed memory scalability of StarNEig when reordering a Schur form.
 Each node contains 28 Intel Xeon E5-2690v4 cores. 
 }
 \label{fig:sep_reorder_scalability}
\end{figure}

Figures \ref{fig:sep_schur_scalability} and \ref{fig:sep_reorder_scalability} give some idea of how well the library is expected to scale in distributed memory.
We note that StarNEig scales reasonably when computing the Schur form and almost linearly when reordering the Schur form.
The iterative nature of the QR algorithm makes the Schur reduction results less predictable because different matrices require different number of bulge chasing steps.
That is, with some matrices, the algorithms ends up performing more consecutive AED steps between the bulge chasing steps, thus leading to faster convergence rate but weaker scalability.

\begin{figure}[h]
 \centering
 \includegraphics[scale=0.75]{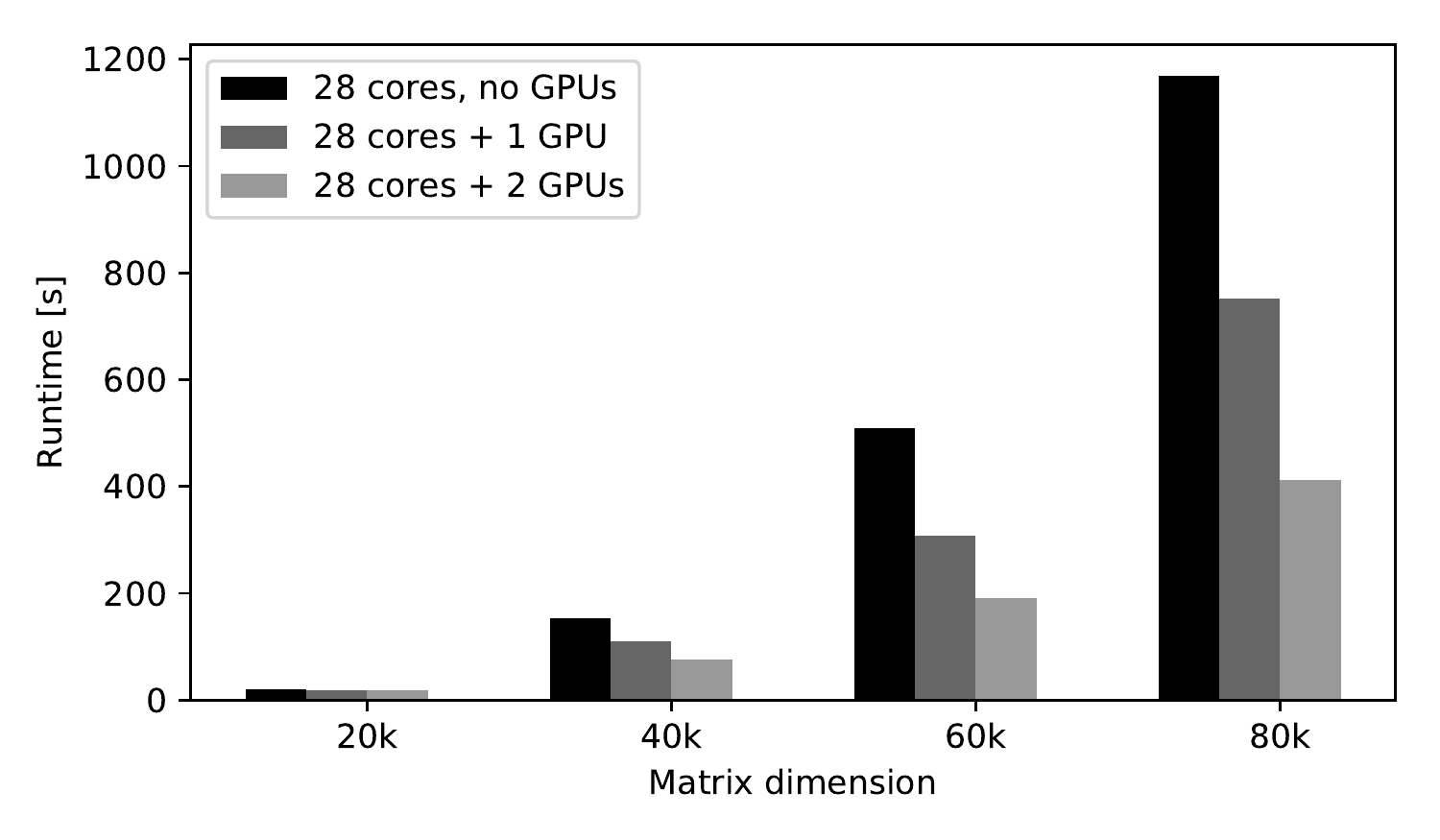}
 \caption{
 GPU performance of StarNEig when reordering a Schur form.
 Each socket (14 Intel Xeon Gold 6132 cores) is connected to one NVIDIA Tesla V100 GPU.
 }
 \label{fig:sep_reorder_gpu}
\end{figure}

Figure \ref{fig:sep_reorder_gpu} demonstrates that StarNEig can indeed take advantage of the available GPUs as long as the matrices are reasonably large.
The introduction of a single V100 GPU gives a speedup of up to 1.65 and the introduction of two V100 GPUs gives a speedup of up to 2.84.
In the best case, the speedup when moving from single V100 GPU to two V100 GPUs is 1.83. 

\begin{table}[h]
\centering
\caption{
 A run time comparison (in seconds) between MKL-LAPACK and StarNEig when computing 35\% of the eigenvectors.
 Run times that are longer than four hours are not tabulated.
} 
\label{tab:sep_eigenvectors_035}
\begin{tabular}{r | c c c c c | c c c c c}
        & \multicolumn{5}{c|}{LAPACK} & \multicolumn{5}{c}{StarNEig} \\
$n$~\textbackslash~cores & 1 & 16 & 32 & 48 & 64 & 1 & 16 & 32 & 48 & 64 \\
\hline
 10\,000 & 95  & 80  & 80 & 80 & 80 & 28 & 2.3 & 1.4 & 1.1 & 1.0 \\
 20\,000 & 784 & 672 & 694 & 673 & 668 & 194 & 15 & 8.1 & 5.9 & 5.3  \\
 40\,000 & 6731 & 5826 & 5681 & 5608 & 5743 & 1364 & 98 & 53 & 38 & 30 \\
 60\,000 & --- & --- & --- & --- & --- & 4231 & 315 & 167 & 116 & 90 \\
 80\,000 & --- & --- & --- & --- & --- & 9221 & 693 & 359 & 252 & 194\\
100\,000 & --- & --- & --- & --- & --- & --- & 1325 & 698 & 483 & 369 \\
120\,000 & --- & --- & --- & --- & --- & --- & 2226 & 1194 & 833 & 647
\end{tabular}
\end{table}

Table \ref{tab:sep_eigenvectors_035} shows how the eigenvector computation routine in StarNEig compares against LAPACK (with parallel BLAS).
In all experiments, 35\% of the eigenvalues were randomly selected.
In single core experiments, StarNEig is between 3.4 and 4.9 times faster than LAPACK. 
This demonstrates demonstrate how efficient the tiled approach is compared to the scalar implementation that exists in LAPACK.
Furthermore, the multi-core experiments demonstrate the poor scalability of the LAPACK implementation.
In particular, the initial solve phase (\ref{eq:evec_solve}) is executed sequentially and the back transformation phase (\ref{eq:evec_backtransform}), which is executed in parallel, constitutes only a small fraction of the total execution time.
The implementation in StarNEig, on the other hand, demonstrates a very good scalability as for the larger matrices ($n \geq 40\,000$) the parallel efficiency stays above 70\%.
In the most extreme case StarNEig is over 190 times faster than LAPACK.

\section{Summary} \label{sec:summary}

This paper presented a new library called StarNEig.
StarNEig aims to provide a complete task-based software stack for solving dense nonsymmetric standard and generalized eigenvalue problems.
StarNEig support both shared and distributed memory machines and some routines in the library can take advantage of the available GPUs.
The paper is mainly aimed at potential users of the library.
Various design choices were explained and contrasted to existing software.
In particular, users who are already familiar with ScaLAPACK should know the following:
\begin{itemize}
 \item StarNEig expect that the matrices are distributed in relatively large blocks compared to ScaLAPACK.
 \item StarNEig should be used in a \textit{one process per node} (1ppn) configuration as opposed to a \textit{one process per core} (1ppc) configuration which is common with ScaLAPACK.
 \item StarNEig implements a ScaLAPACK compatibility layer.
\end{itemize}

The performance of the library was demonstrated with a set of computational experiments.
The presented results show the following:
In the Hessenberg reduction phase, StarNEig is competitive with LAPACK, ScaLAPACK (in single node) and MAGMA (single GPU).
In the Schur reduction phase, StarNEig is between 1.6 and 2.9 times faster than ScaLAPACK.
In the eigenvector computation phase, StarNEig's parallel and robust implementation significantly outperforms LAPACK in both single-core and multi-core settings with recorded speedups as large as 190.
In the eigenvalue reordering phase, StarNEig is between 2.8 and 5.0 times faster than ScaLAPACK and scales nearly linearly.

The library is still incomplete.
Future work with StarNEig includes the implementation and integration of the missing software components.
Support for complex valued matrices is also planned.
The GPU support, and the multi-GPU support in particular, are still under active development.
The authors hope to start a discussion which would help guide and prioritize the future development of the library.

\section*{Acknowledgments} 
\label{sec:acknowledgements}

StarNEig has been developed by Mirko Myllykoski (who has designed and implemented the Hessenberg reduction, Schur reduction and eigenvalue reordering phases), Carl Christian Kjelgaard Mikkelsen (who has designed and implemented the eigenvector computation phase for generalized eigenvectors), Angelika Schwarz (who has designed and implemented the eigenvector computation phase for standard eigenvectors), Lars Karlsson (who has worked with the underlying theory behind the robust computation of eigenvectors), and Bo K{\aa}gstr{\"o}m (who was the coordinator and scientific director of the NLAFET project).
This work is part of a project (NLAFET) that has received funding from the European Union's Horizon 2020 research and innovation programme under grant agreement No 671633. 
This work was supported by the Swedish strategic research programme eSSENCE and Swedish Research Council (VR) under Grant E0485301.
We thank the High Performance Computing Center North (HPC2N) at Ume{\aa} University for providing computational resources and valuable support during test and performance runs.
The authors acknowledge the work of Mahmoud Eljammaly and Bj\"{o}rn Adlerborn during the NLAFET project.
The authors would also like to thank the StarPU team at INRIA for rapidly providing good answers to several questions related to StarPU.
%Finally, the author thanks the anonymous reviewers for their valuable feedback.

%\FloatBarrier

%\nocite{*}% Show all bib entries - both cited and uncited; comment this line to view only cited bib entries;
\bibliography{myllykoski}

\end{document}